\documentclass[aps,physrev,twocolumn,superscriptaddress]{revtex4-2}

\usepackage{amsmath,amssymb,amsfonts}
\usepackage{graphicx}
\usepackage{xcolor}
\usepackage{hyperref}

\begin{document}

\title{Hydrodynamic modeling of liquid metal field emitter deformation under combined RF and DC magnetic fields}

\author{Yuan Zhi Lim}

\affiliation{ESPCI-Paris, PSL Research University, 75005, Paris, France}

\author{Juliette Plouin}
\email[]{juliette.plouin@cea.fr}

\affiliation{CEA, Université Paris-Saclay, 91191, Gif-sur-Yvette, France}

\author{Guillaume Ferrand}

\affiliation{CEA, Université Paris-Saclay, 91191, Gif-sur-Yvette, France}

\date{\today}

\begin{abstract}
Radio-frequency (RF) breakdowns, also known as vacuum (or electrical) arcs and breakdowns, or electrical discharges, have been studied for over decades, but the mechanisms of the RF breakdown are still questionable. This phenomenon reduces the stability of the beam, damages the RF cavities, and thus limits the achievable accelerating field in the particle accelerator. In the case of Muon Collider, the RF cavities have to withstand an electric field of the order of multi tens of MV/m and a DC magnetic field of a few tens of teslas, with a minimum rate of RF breakdown. Different models have been proposed, and some important factors that could lead to the RF breakdown have been identified, such as the field emission current, irregularity of the surface , presence of hot particles on the surface and the role of the DC magnetic field. In this paper, we give insights to the mechanism of RF breakdowns using a hydrodynamic  and a magnetohydrodynamic approach, focusing on understanding the forces that deform the liquid copper emitter. We found that if the local RF electric field is sufficiently high, it could provoke the emission of hot copper droplets into the cavity. Our simulations use the computational fluid dynamics module, the magnetic and electric field module, and the level set method in the COMSOL software to simulate the behavior of a copper liquid emitter with the Navier-Stokes equation under the influence of the pressure radiation, and the surface tension force.
\end{abstract}

\maketitle

\section{Introduction}
Muon colliders have a great potential for high-energy physics. They can offer collisions of point-like particles at very high energies, since muons can be accelerated in a ring without suffering from synchrotron radiation losses. The concept of using colliding beam with positive and negative muons dates back to the late 60th \cite{tikhonin2008}, but encountered the difficult challenge of producing bright muon beams within the short life time of muons \cite{long_muon_2021}. The principle of muon beam generation is the following: a proton beam is  directed onto a target, where positive and negative pions are produced that subsequently decay into positive and negative muons accompanied by neutrinos and antineutrinos. The oppositely charged muons are separated into two beams that are then cooled using the ionization-cooling technique and afterwards accelerated in several stages before they are injected into the collider ring. The cooling system is essential to reduce the large initial phase space by a factor of more than five orders of magnitude to that needed for a practical collider. Without such cooling, the luminosity would be several orders of magnitude too small \cite{long_muon_2021}. 

The cooling system reduces both transverse and longitudinal momentum of the beam, and requires the presence of radiofrequency (RF) cavities to restore the  longitudinal momentum and preserve beam acceleration.

In the 2000's, the possibility of performing this so-called 6D cooling of muons has been demonstrated by the Muon Ionisation Cooling Experiment (MICE)
\cite{MICE_2020}, and a US Muon Accelerator Program (MAP) has been developed to carry out the R\&D necessary to evaluate the feasibility of future facilities based on muon accelerators \cite{Palmer_2013}. 

In 2020, a new international design study for a future muon collider has been initiated, with the launch of the European MuCol program \cite{MuCol} and of the International Muon Collider Collaboration (IMCC) \cite{IMCC} dedicated to overcome the technical challenges and prepare a design study \cite{long_muon_2021}.

The state of the work on accelerator design and technology within this collaboration has submitted an input to the 2026 European Strategy for Particle Physics Update (ESPPU) \cite{ESPPU_2026}. The rectilinear 6D cooling system required to reduce the beam emittance is divided in two sections, placed before (A-type) and after (B-type) a bunch merge system. The pre- and post- bunch merge sections are respectively divided in four and ten stages, and each stage contains some tens of cooling cells \cite{Zhu_2025}.

\begin{figure}[htb]
    \centering
    \includegraphics[width=0.45\textwidth]{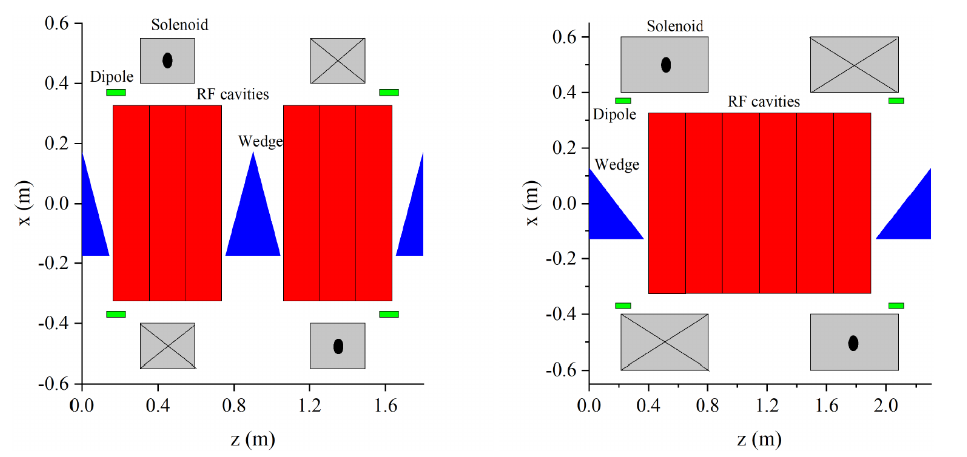}
    \caption{Schematic of the A-type (left) and B-type (right) cooling cell layout. Figure from \cite{ESPPU_2026}.}.
    \label{fig:Coolcell}
\end{figure}

As detailed in Fig. \ref{fig:Coolcell}, each of the cooling cells contains wedge-shaped absorbers (6D cooling), RF cavities (longitudinal momentum restoring), solenoids and dipoles (beam focusing). The field levels required in each cell is of the order of multi tens of MV/m for cavities and multi-T for solenoids. The cavity frequencies are 352 or 704~MHz.

Due to the short lifetime of muons at rest, the cooling cells must be compact, resulting in the coils and cavities being very close together. This means that the RF cavities must operate in a multi-T magnetic field environment. This dismisses the use of superconducting cavities, and requires normal conducting materials like copper. Moreover, experimental tests carried out in the early 2000's have shown a decrease of the RF breakdown electric field threshold in presence of such magnetic fields, as well as an increase of the BD rate \cite{norem_dark_2003,moretti_effects_2005,palmer_rf_2009,bowring_operation_2020}.
In consequence, requiring such RF fields in multi-T environment is one of the challenge of the cooling cells, and requires a dedicated R\&D programme to ensure the feasibility of such fields  with a reasonable breakdown rate.

In the past decades, and associated to experimental tests, theoretical and modeling approaches have been developed to explain this phenomenon, none of them being entirely conclusive. 

Norem et al. made exhaustive Fowler Nordheim (FN) current measurements with a 6 cells 805~MHz cavity \cite{norem_dark_2003}. In presence of B field, both dark current and level of radiation are increased, and post mortem observations showed concentric bands of burned spots on the cavity faces. The authors assumed those were associated to the hit of emitted electrons, accelerated from their emission point by $\vec{E}\times\vec{B}$ drifting. They also assume a direct effect on the emitter itself due to the Hall effect.

Palmer and al. propose a "beamlet" model for explaining the RF breakdown increase in presence of magnetic field \cite{palmer_rf_2009}. Dark current electrons are emitted from an asperity, accelerated by the RF fields, and impact a cavity wall. With sufficient magnetic field, the emitted electrons are focused to
small spots, where they heat the surface sufficiently to cause damage. This model has been developed in great detail in \cite{bowring_operation_2020} where the beamlet radius is calculated against the magnetic field; the deposited power and the temperature increase are calculated. This latter is compared to a temperature threshold above which plastic deformation and surface damage may affect cavity behavior. 

These studies contribute to the in-depth study of the breakdown phenomena in the presence of multi~tesla DC magnetic field. The models proposed in \cite{norem_dark_2003} and \cite{moretti_effects_2005} focus on the B field effect on the emitter itself, while \cite{palmer_rf_2009} and \cite{bowring_operation_2020} focus on the effect on the emitted electrons trajectories. 

The necessity of gathering new experimental and fundamental studies is highlighted in the new programs for a future muon collider. In particular, experimental setups shall be developed to measure the effect of numerous parameters in the DC magnetic field influence on the breakdown: magnetic field orientation and position, RF frequency, cavity material and operating temperature \cite{nanni_2025}\cite{plouin_2025}. The resulting experimental results will allow benchmarking against theory and modeling tools.
 
Obviously, the study of general breakdown phenomena (without B field) is very exhaustive \cite{wuensch_fundamentals_2025}. In particular, the work developed by Abe et al. \cite{abe_2018,abe_2019,abe_2025} caught our attention and influenced our choice of the model presented in this paper. They made direct in-situ observation inside a 508.9~MHz cavity with mirrors and high speed/hyperspectral cameras. Not only they could observe bright spots on the cavity surface with a size of approximately 0.1 mm or smaller, but they were able to observe bright objects flying inside the cavity. These flying objects, moving with a velocity around 10~m/s, finally impact the cavity surface and generate a vacuum breakdown. They found that bright spots on the cavity surface or flying bright objects inside the cavity are the dominant causes of breakdown of the 508.9~MHz cavity. 

The observation of cone-like structures at the metal surface of cavity after breakdown has shown to be compatible with the hypothesis that conducting liquids subjected to fields tend to form cone-shaped structures (Taylor cones) \cite{antoine_erratum_2012}, stabilized by the balance of electrostatic and surface tensions forces. If the field is increased, the cones are stretched, become unstable and the emission of micro-droplets is likely to start \cite{taylor_disintegration_1964}. 

These studies have convinced us to develop a fluid dynamics model based on the forces generated by the RF field, the magnetic field, and the forces inherent to the fluid. Actually, in the physics of hydrodynamics, not only the dynamic of a conducting liquid metal tip can be described by the RF radiation pressure and by the surface tension, but also the effect of the DC magnetic field is described in the magnetohydrodynamics model \cite{alfven_1942}\cite{davidson2017}.

The development of this model is presented in this paper, and  represents our contribution to understanding the phenomena underlying the effect of magnetic fields on breakdown.

\section{Description of the hydrodynamic (HD) model\label{des_hd}}

\begin{figure}[htb]
    \centering
    \includegraphics[width=0.45\textwidth]{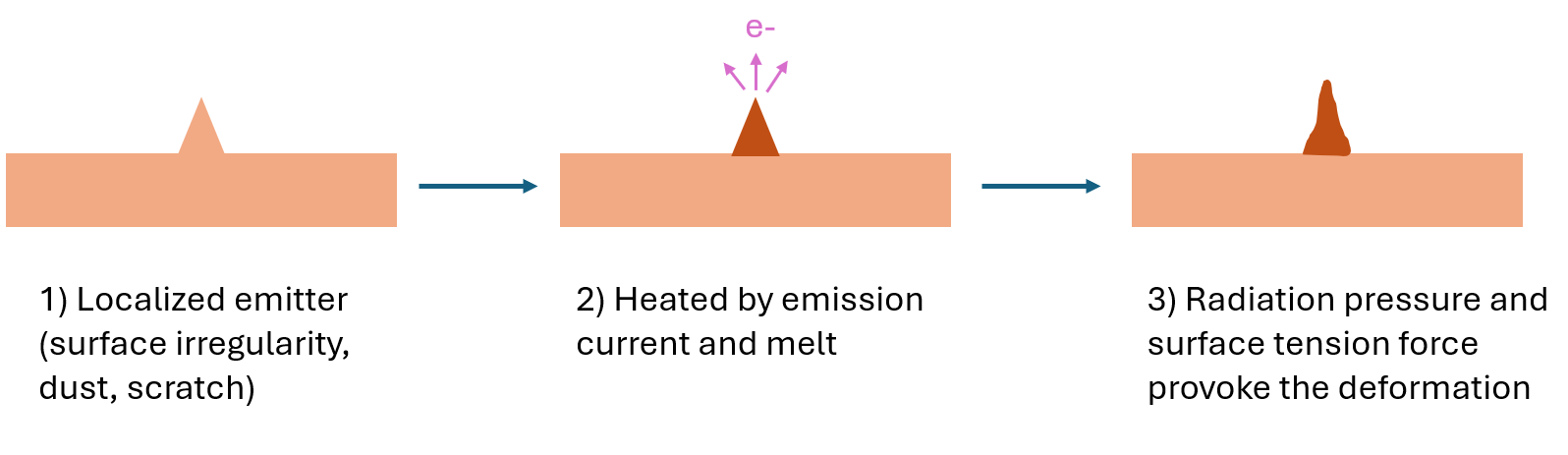}
    \caption{Mechanism of the breakdown process.}
    \label{fig:Mechanism}
\end{figure}

The RF breakdown (Fig.~\ref{fig:Mechanism}) is initiated by a local defect (surface irregularity, dust, scratch). The local defect provokes a huge emission current due to a large field enhancement factor ($\beta = 10 \sim 300$ \cite{antoine_erratum_2012}) at the apex of the field emitter, sufficient to heat and melt the local defect. The RF electromagnetic force (radiation pressure) and the surface tension force compete with each other and provoke deformation, causing the possible ejection of small parts of the liquid emitter from the surface of the cavity.

\begin{figure}[htb]
    \centering
    \includegraphics[width=0.45\textwidth]{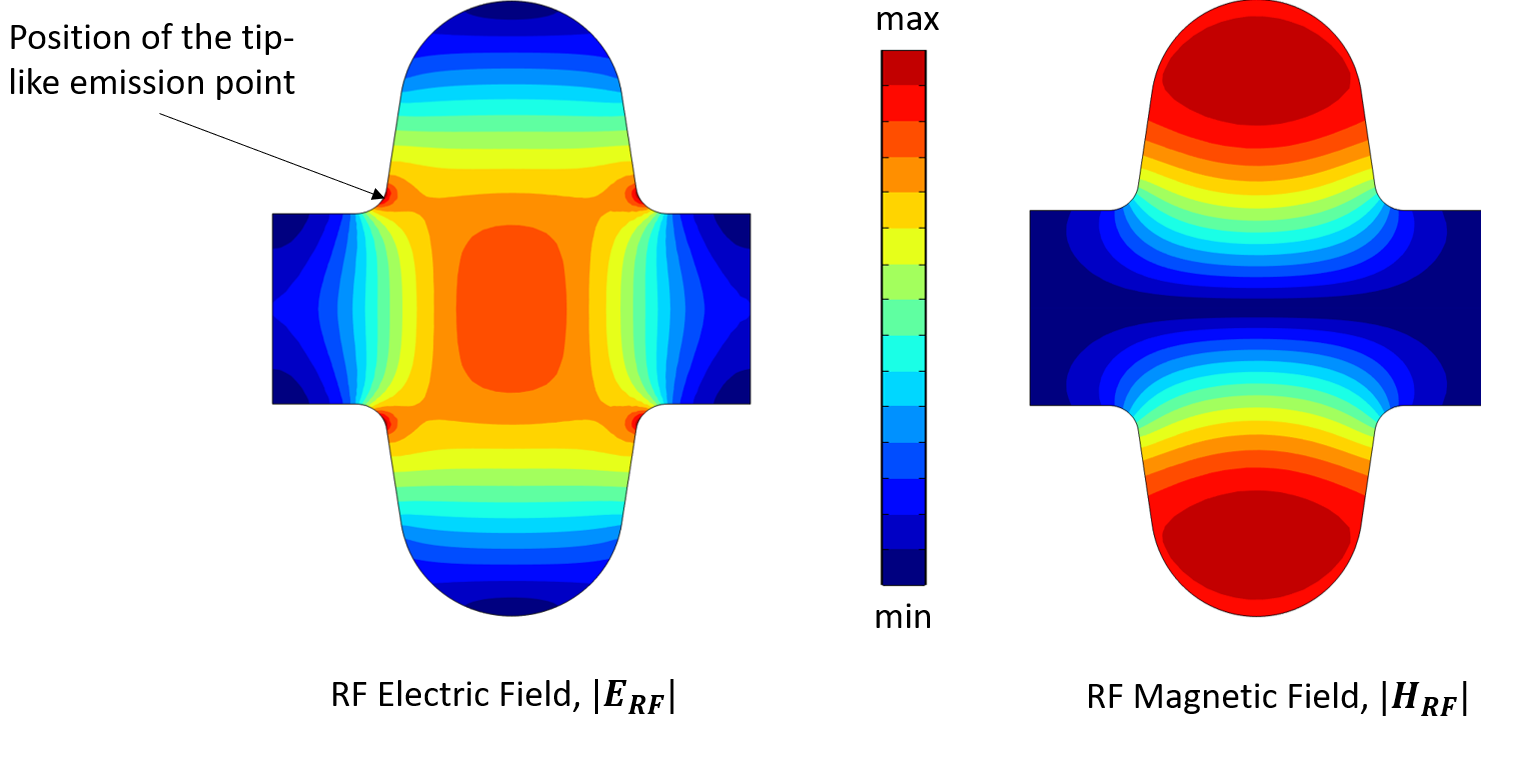}
    \caption{A 704 MHz elliptical cavity showing the possible location for a tip-like emission point.}
    \label{fig:Cavity}
\end{figure}

 \begin{figure}[htb]
    \centering
    \includegraphics[width=0.45\textwidth]{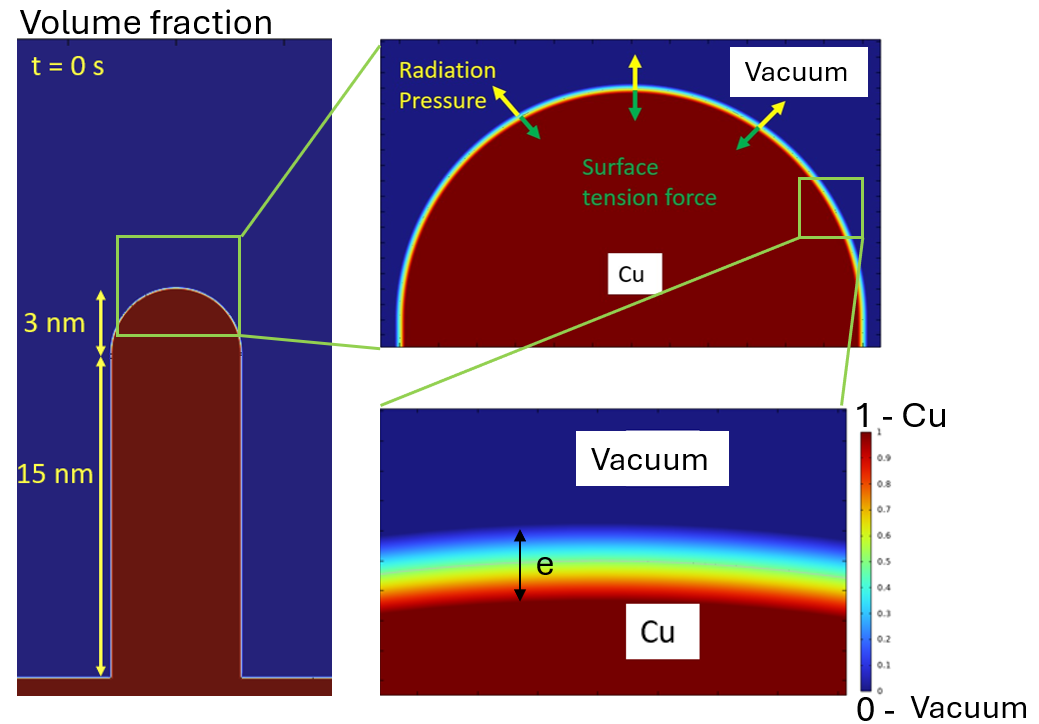}
    \caption{A tip-like emission point on the surface of the cavity. The radiation pressure pulls the liquid copper towards vacuum and the surface tension force opposes the radiation pressure. The colorbar indicates the volume fraction of copper, from 0 (vacuum) to 1 (copper) The liquid copper/vacuum interface is described by a gradient of the volume fraction across a thin layer ($e$), governed by the level set module in COMSOL.}
    \label{fig:Cu Metal Tip}
\end{figure}

In our study, we consider a nanometer-long tip-shaped emission point located on a high RF electric field but low RF magnetic field area on the surface of the cavity (Fig.~\ref{fig:Cavity}), made of copper. Due to a high local RF electric field, the emission current causes the copper tip to heat up and thus melt. Our model considers that the heating and melting process had already occurred and that the copper emission point is molten at 1100°C. The liquid copper becomes a conductive fluid subjected to RF electromagnetic fields. At the liquid/vacuum interface, two pressures perpendicular to the interface are competing with each other (Fig.~\ref{fig:Cu Metal Tip}). The radiation pressure pulls the liquid copper towards the vacuum, and the surface tension force opposes the radiation pressure. This leads to deformations, depending on the dominant component.

\subsection{Surface forces} \label{subsection:Surface forces}

The deformations of the liquid copper depend on two surface forces: the radiation pressure and the surface tension force, as well as a body force due to viscosity described in Section~\ref{subsection: Hydrodynamic equation}. 

The radiation pressure is the time-average of the RF electromagnetic force and is given by the following formula \cite{Prad}:

\begin{equation} \label{eqn:Pression de radiation}
    P_{rad} = \frac{1}{4}\left(\varepsilon_0 |\mathbf{E_{RF}}|^2 - \mu_0 |\mathbf{H_{RF}}|^2 \right) \ [N/m^2]
\end{equation}

where $|\mathbf{E_{RF}}|$ and $|\mathbf{H_{RF}}|$ are the cavity RF electric and magnetic fields at the surface of the conducting copper. Note that this formula has been demonstrated when the skin-depth remains small compared to the conductor dimension, which will be discussed later for this model. When $P_{rad} > 0$, the radiation pressure pulls the liquid copper towards the vacuum. Since RF breakdown is mostly observed in the high $|\mathbf{E_{RF}}|$ cavity field region, we neglect the contribution from the magnetic field $|\mathbf{H_{RF}}|$. We take into account the field enhancement factor $\beta$, due to an increase in the local electric RF field caused by geometric effects of the field emitter. Equation \ref{eqn:Pression de radiation} turns into:

\begin{equation} \label{eqn:Prad}
    P_{rad} = \frac{1}{4} \varepsilon_0 \beta^2 |\mathbf{E_{RF}}|^2 \ [N/m^2]
\end{equation}

The surface tension force (Laplace pressure) is the pressure difference between the inside and the outside of a curved surface caused by the surface tension. The surface tension force is described as follows:

\begin{equation}\label{eqn:Fst}
F_{st} = \gamma_{st} \kappa  \ [N/m^2]
\end{equation}

where $\gamma_{st}$ is the surface tension coefficient between liquid copper and vacuum, and $\kappa$ is the local curvature. 

\subsection{The numerical Dirac function and body forces}

The Navier-Stokes equation is used to simulate the deformations of the liquid copper in COMSOL. Instead of using the surface forces described in Section~\ref{subsection:Surface forces}, we need to first transform the surface forces into body forces and then inject them into the Navier-Stokes equation. We introduce $\delta$, the numerical Dirac delta function that acts only on a thin layer ($e$) around the interface as shown in Fig.~\ref{fig:Cu Metal Tip}, and is defined as follows:

\[
\delta(\mathbf{r}) \ [1/m] 
\begin{cases}
= 6 |\nabla \phi| |\phi(1-\phi)| \neq 0, & \text{at the interface},\\[4pt]
= 0, & \text{elsewhere}
\end{cases}
\]

with

\[
\int \delta dl = 1
\]
The  $\phi$ coefficient is the volume fraction of vacuum.

The two surface forces are converted into body forces by multiplying with the numerical Dirac function. The numerical Dirac function is governed by the level set module in COMSOL \cite{comsol1,comsol2}.

\subsection{Hydrodynamic equation} \label{subsection: Hydrodynamic equation}
The movements of liquid copper is described by the Navier-Stokes equation, under the influence of the radiation pressure and the surface tension force. The generalized Navier-Stokes equation used in COMSOL simulations is given by Equation~\ref{eqn:Generalised NS}.

\begin{equation} \label{eqn:Generalised NS}
\rho
\left(
\frac{\partial \mathbf{u}}{\partial t}
+
(\mathbf{u} \cdot \nabla)\mathbf{u}
\right)
=
-\nabla \cdot (p \mathbf{I} + \mathbf{K})
+
\mathbf{F}
+
\rho \mathbf{g} \ \ [N/m^3]
\end{equation}

with $\rho$ the density of the fluids, $\mathbf{u}$ the velocity of the fluids, $p$ the pressure, $\mathbf{I}$ the identity matrix, $\mathbf{K}$ the viscous stress tensor, $\mathbf{F}$ the external forces and $\mathbf{g}$ the gravitational acceleration.

In our case, we assume that the liquid copper is an incompressible Newtonian fluid and there is no gravity acting on our system. By injecting the radiation pressure and the surface tension force into Equation~\ref{eqn:Generalised NS}, we obtain Equation~\ref{eqn:HD}, which is the governing equation used in COMSOL simulations.

\begin{equation} \label{eqn:HD}
\begin{aligned}
\rho
\left(
\frac{\partial \mathbf{u}}{\partial t}
+
\underbrace{
(\mathbf{u} \cdot \nabla)\mathbf{u}
}_\text{convection}
\right)
&=
\underbrace{-\nabla p}_{\substack{\text{pressure} \\ \text{gradient}}}
+
\underbrace{\eta \nabla^2 \mathbf{u}}_\text{viscous term} \\
&\qquad +
\underbrace{P_{rad} \delta \cdot \mathbf{n}}_{\substack{\text{radiation} \\ \text{pressure}}}
+
\underbrace{F_{st} \delta \cdot \mathbf{n}}_{\substack{\text{surface} \\ \text{tension force}}} \ [N/m^3]
\end{aligned}
\end{equation}

where $\eta$ is the dynamic viscosity, $\delta$ is the numerical Dirac delta function and $\mathbf{n}$ is oriented outwards the copper.

\subsection{Threshold to have a possible breakdown event\label{subsec:threshold}}

From Equation~\ref{eqn:HD}, we can deduce a theoretical threshold for a possible breakdown event to occur. First, there is no pressure gradient due to the incompressibility of liquid copper, so $\nabla p$ is zero. Then, at the start of the simulation (t~=~0 s), the velocity $\mathbf{u}$ is zero everywhere, and the velocity direction just after t~=~0 depends on the sign of $\partial \mathbf{u}/\partial t$ and thus on the sign of $P_{rad} + F_{st}$. Using Equation~\ref{eqn:Prad} and \ref{eqn:Fst}, the threshold for the tip being pulled by the radiation pressure, and for a possible breakdown event to occur is $P_{rad} > F_{st}$, which leads to:

\begin{equation} \label{eqn:threshold}
    |\mathbf{E}_{RF}| \;>\; \frac{2}{\beta} \sqrt{\frac{\gamma_{st}\,\kappa}{\varepsilon_0}}
\end{equation}


\section{Simulations in COMSOL Multiphysics\label{sec:Comsol}}

 \begin{figure}[htb]
    \centering
    \includegraphics[width=0.2\textwidth]{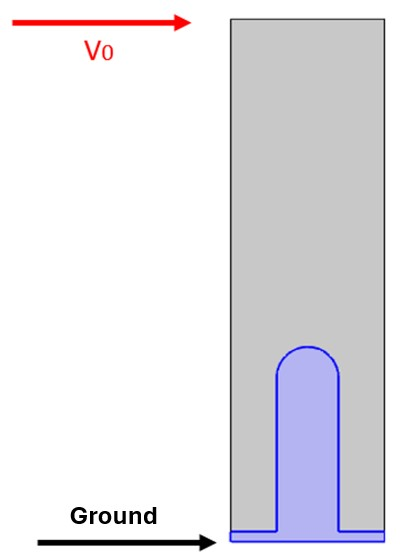
}
    \includegraphics[width=0.35\textwidth]{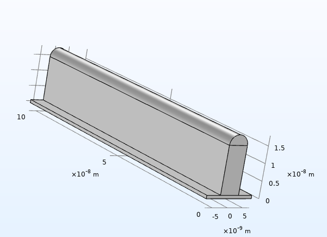}
    \caption{A 2D liquid copper emitter (blue) inside a simulation box filled with vacuum (grey) and the 3D view (projection) of the tip.}
    \label{fig:Cu Tip}
\end{figure}

\begin{figure}[htb]
    \centering
    \includegraphics[width=0.45\textwidth]{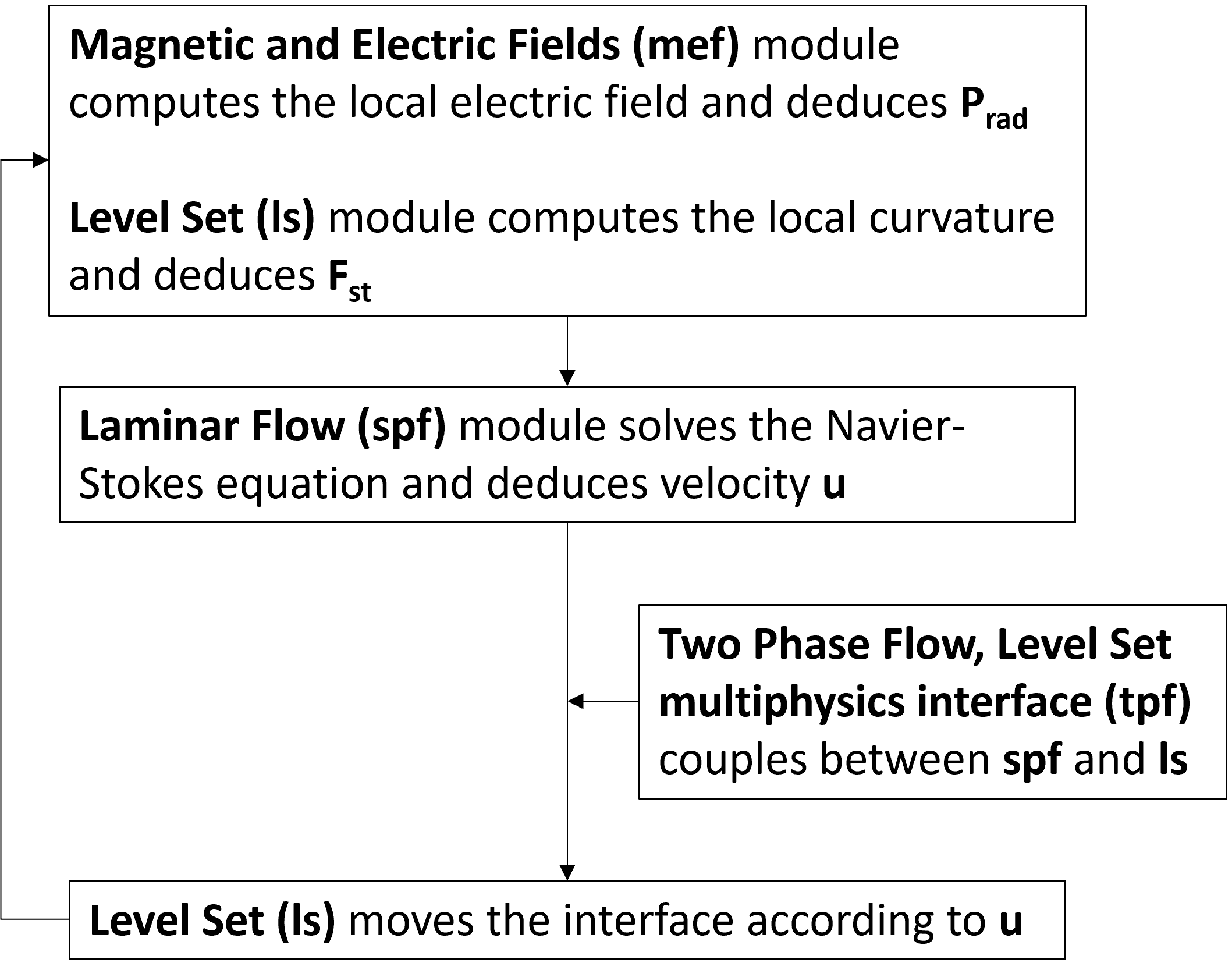}
    \caption{COMSOL Setup for our 2D model}
    \label{fig:COMSOL}
\end{figure}

We developed a model in COMSOL to simulate the deformation of a liquid copper emitter in presence of an electric field and under the guidance of the Navier-Stokes (Eq.~\ref{eqn:HD}). The computation setup, described hereafter, consists into a time loop calculation involving several multiphysics solvers and interfaces. It is time-consuming for a 3D resolution, so we used a 2D model, with a tip shaped domain (Fig.~\ref{fig:COMSOL}, top), considered to be invariant in the orthogonal direction  (Fig.~\ref{fig:COMSOL}, bottom). This is a strong simplification of the model, however it allows relevant results for the liquid copper emitter behavior. Moreover, due to computing time limitation, the complete simulation domain is limited in the horizontal direction. The electric field lines are constraints by the proximity of the vertical boundary positions, resulting in an under evaluated value for the enhancement factor $\beta$ (see Sec.~\ref{sec:hd}). Finally, the simulation in COMSOL could not converge with the vacuum conductivity equal to zero. We set its conductivity to 1~S/m, after having verified that the results did not change by varying the conductivity around this value.

Since only the amplitude of the RF field counts as we used the time-average formula, we fixed a voltage difference across the two ends of the simulation box  ($V_0$ in Fig.~\ref{fig:Cu Tip}) to impose a constant electric field $|\mathbf{E_{RF}}|$ using the magnetic and electric interface ($mef$). The $mef$ module in COMSOL computes the local electric field, while the laminar flow interface ($spf$) computes the velocity of the liquid copper emitter under the guidance of the Equation~\ref{eqn:HD}. The level set ($ls$) method in COMSOL is used to compute the local curvature, describe the movements of the liquid copper/vacuum interface by using the advection equation, and ensure the numerical stability of the interface \cite{comsol1,comsol2}. The two phase flow, level set multiphysics interface couples between $spf$ and $ls$ by injecting the velocity solved in $spf$ into $ls$. Initially, COMSOL computes the local electric field $|\mathbf{E_{RF \ loc}}| = \beta|\mathbf{E_{RF}}|$ applied to the liquid copper emitter using $mef$, and the local curvature is deduced by the level set. This generates the radiation pressure and the surface tension force, thus causing the interface to start deforming. At the next time step, COMSOL recomputes the local electric field and the local curvature based on its new shape and continues to deform the copper emitter. The diagram in (Fig.~\ref{fig:COMSOL}) shows the setup and the time-loop used in COMSOL simulations.

\section{Result of the Hydrodynamic (HD) model\label{sec:hd}}

 \begin{figure}[htb]
    \centering
    \includegraphics[width=0.45\textwidth]{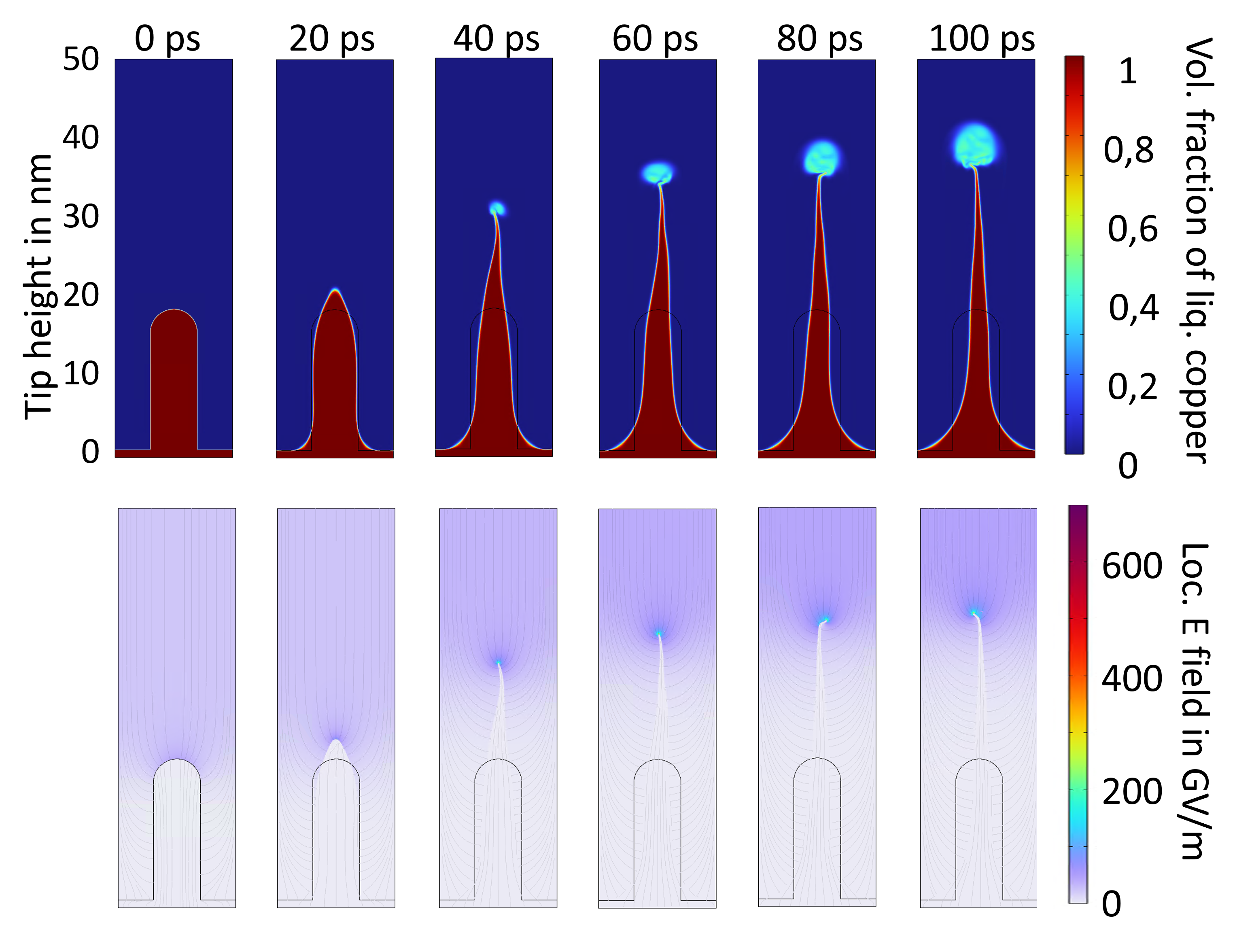}
    \caption{Deformation of the liquid copper emitter (top) and local electric field map (bottom) at different times.}
    \label{fig:HD Model}
\end{figure}

\begin{figure}[htb]
    \centering
    \includegraphics[width=0.45\textwidth]{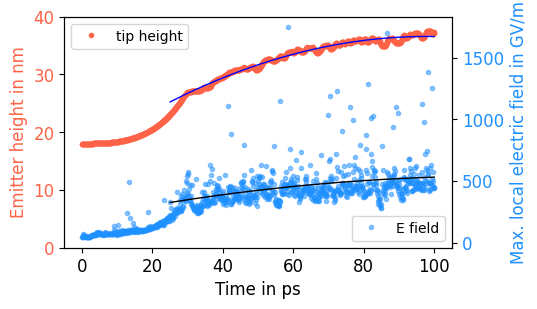}
    \caption{Variation of the emitter height (red) and the local electric field (blue) as a function of time.}
    \label{fig:Tip Elongation}
\end{figure}

We simulated the deformation of the liquid copper emitter with an initial height of 18~nm and a radius of 3~nm (Fig.~\ref{fig:HD Model}, t=0). The  surface tension coefficient and the copper viscosity, considered to be constant, are $\gamma_{st}$~=~1.3~N/m and $\eta$~=~$4.03~10^{-3}~N.s/m^2$.  The liquid copper conductivity has been fixed to $4.8~10^6~S/m$, its value at 1100°C.  We applied a field of $|\mathbf{E_{RF}}|$ = 15~GV/m, resulting in an initial local electric field $|\mathbf{E_{RF \ loc}}|$ = 42~GV/m at the tip of the liquid copper emitter. The corresponding $\beta$ value, equal to 2.8, lower than expected for such tip pattern, as discussed in Section~\ref{sec:Comsol} . Fig.~\ref{fig:HD Model} shows how the liquid copper emitter is pulled by the radiation pressure and how the local electric field increases as the radius of the tip gradually decreases. The graph in Fig.~\ref{fig:Tip Elongation} shows the increase in height of the tip and the local field enhancement as a function of time. The height estimate becomes less accurate after 30~ps, due to the appearance of a region where the volume fraction of copper is in between 0 and 1 (cyan spot in Fig.~\ref{fig:HD Model}): the boundary between vacuum and copper is numerically represented by a non physical mixture of fluids.

The electric field in Fig.~\ref{fig:Tip Elongation} is calculated at each time step as the maximal field in the simulation domain, situated close to the tip apex. Since the region of electric field concentration decreases over time, and reaches the size of a mesh cell, the graph tends to exhibit numerical “hot spots”. However, as for the height evolution, a general trend becomes apparent. 

We first observe a fast increase of both quantities when the tip begins to stretch and becomes thinner, due to the formation of a positive feedback loop. After 30~ps, a part of the copper escapes from the tip. This can be interpreted as the ejection of hot copper particles into the vacuum. Then, the overall shape of the emitter keeps the same, and we can assume the establishment of a balance between the radiation pressure and the surface tension. The increase of both height and field enhancement factor becomes slower.

In Section~\ref{des_hd}.A, we mentioned that the radiation pressure formula was demonstrated for a skin-depth small compared to the conductor dimensions. Actually, with a typical frequency of 704~MHz, the skin-depth is 8.6~µm, three order of magnitude over the tip dimensions (height: 18~nm, radius: 3~nm). One could consider that the definition of the radiation pressure in our model is not appropriate. However, it will be shown in the following that the tip behavior does not significantly vary:
\begin{itemize}
    \item when the dimensions become larger that the skin depth (height: 180~µm, radius: 30~µm), see Sec.~\ref{sec:mhd}
    \item for the same dimensions (height: 18~nm, radius: 3~nm) but with an atomistic model.(Sec.~\ref{sec:comp}). 
\end{itemize}
It is likely that the definition of the radiation pressure given in Section~\ref{des_hd} gives appropriate results. However, the representation of the surface pressure for geometries much smaller than the skin-depth would deserve further exploration.

\section{Description of the magnetohydrodynamic (MHD) model}

This section aims to understand the effect of the DC magnetic field on the deformation of the liquid copper tip. 

The generalized Ohm's law is used to compute the induced current and the associated induced Lorentz force due to the presence of $\mathbf{B_{DC}}$. 

The generalized Ohm's law formula is given by:
\begin{equation}
    \mathbf{J} =  \underbrace{ \sigma \mathbf{E_{tot}}}_{\text{displacement current}}
           +  \qquad \underbrace{\sigma (  \mathbf{u} \times \mathbf{B_{tot}})}_{\text{induced current}} \ [A/m^2]
\label{eq:ohm}   
\end{equation}

where $\sigma$ represents the fluid conductivity, $\mathbf{u}$ is the velocity of the fluid,

The Lorentz force formula is given by:
\begin{equation} \label{eqn:Lorentz}
    \mathbf{F_L} = \rho_{en} \mathbf{E_{tot}} + \mathbf{J} \times \mathbf{B_{tot}} \ [N/m^3]
\end{equation}  

where $\rho_{en}$ is the charge density that differs from the electro-neutrality. In the case of a non-relativistic magnetohydrodynamic model, $\rho_{en} \mathbf{E_{tot}}$ is negligible compared to $\mathbf{J} \times \mathbf{B_{tot}}$ \cite{mhd_ensta}. 

The total fields are $\mathbf{E_{tot}}~=~\mathbf{E_{RF}}~+~\mathbf{E_{ind}}$ and $\mathbf{B_{tot}}~=~\mathbf{B_{RF}}~+~\mathbf{B_{DC}}~+~\mathbf{B_{ind}}$. The quantities $\mathbf{E_{ind}}$ and $\mathbf{B_{ind}}$ are the induced fields deduced by Maxwell's equations, which will be neglected. Moreover, the tip is situated in an area where the RF magnetic field is very low (see Sec.~\ref{des_hd}.A), so $\mathbf{B_{RF}}~<<~\mathbf{B_{DC}}$.

Equation \ref{eqn:Lorentz} turns into:
\begin{equation} \label{eqn:Lorentz simplified}
\begin{aligned}
    \mathbf{F_L} &= \mathbf{J} \times \mathbf{B_{tot}} \\ &= 
    \sigma \mathbf{E_{RF}} \times \mathbf{B_{DC}}
    + \sigma (\mathbf{u} \times \mathbf{B_{DC}}) \times \mathbf{B_{DC}}
\end{aligned}
\end{equation}

The time-average $<\mathbf{E_{RF}} \times \mathbf{B_{DC}}>$ is ~0 so we cancel this term.
Finally, we get Equation~\ref{eqn:F_L}:

\begin{equation} \label{eqn:F_L}
    \mathbf{F_L} = 
    \underbrace{\sigma (\mathbf{u} \times \mathbf{B_{DC}}) \times \mathbf{B_{DC}}}_{\text{force which opposes the initial movement}} \ [N/m^3]
\end{equation}

By adding this force to the hydrodynamic equation, we arrive at the magnetohydrodynamic equation:

\begin{equation}\label{eqn:MHD}
\begin{aligned}
\rho
\left(
\frac{\partial \mathbf{u}}{\partial t}
+
\underbrace{
(\mathbf{u} \cdot \nabla)\mathbf{u}
}_\text{convection}
\right)
&=
\underbrace{-\nabla p}_{\substack{\text{pressure} \\ \text{gradient}}}
+
\underbrace{\eta \nabla^2 \mathbf{u}}_\text{viscous term}
+
\underbrace{P_{rad} \delta \cdot \mathbf{n}}_{\substack{\text{radiation} \\ \text{pressure}}}
\\ &\quad +
\underbrace{F_{st} \delta \cdot \mathbf{n}}_{\substack{\text{surface} \\ \text{tension force}}}
+
\underbrace{\mathbf{F_L}}_\text{Lorentz force} \ [N/m^3]
\end{aligned}
\end{equation}

\begin{figure}[htb]
    \centering
    \includegraphics[width=0.45\textwidth]{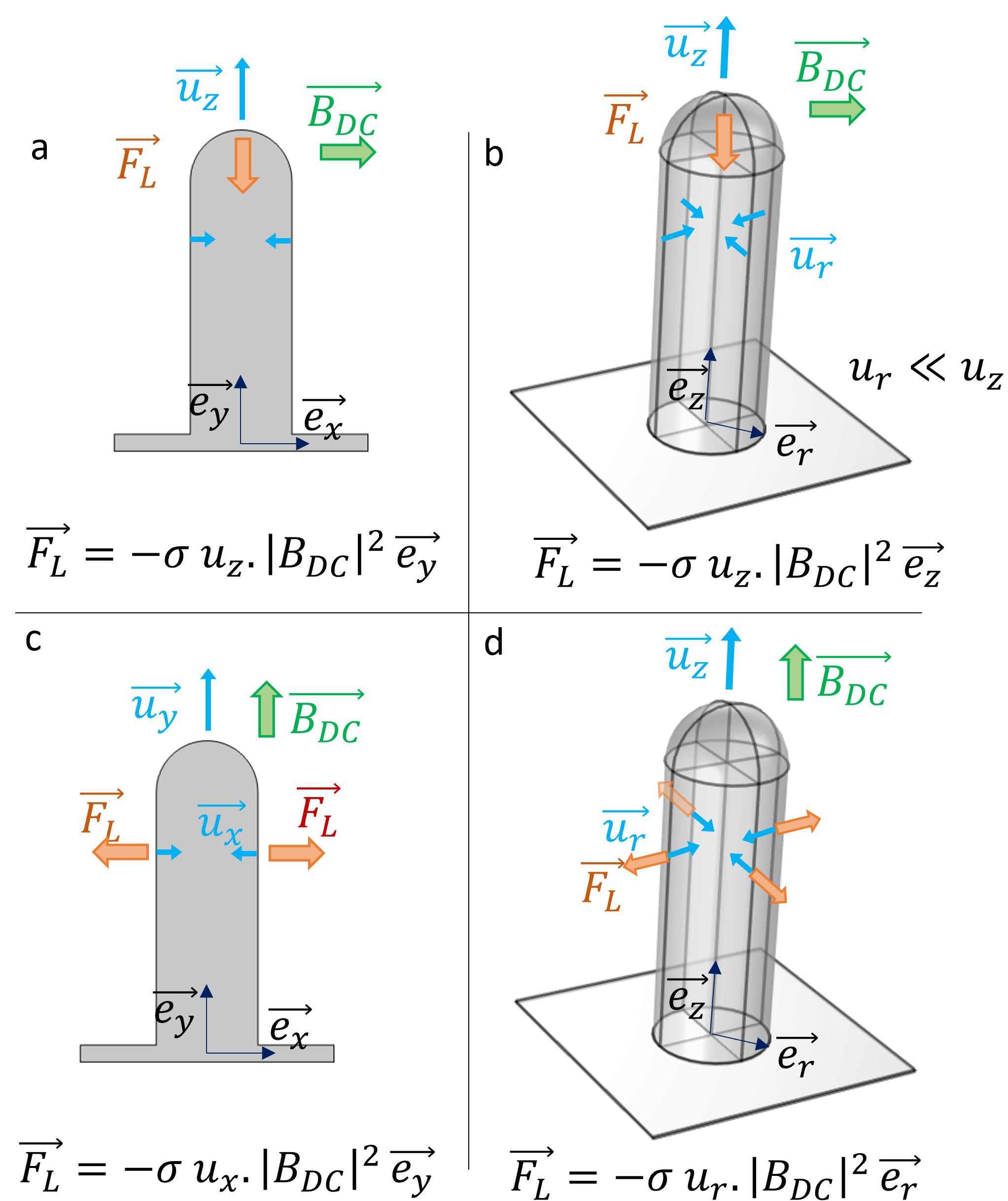}
    \caption{Illustration of the Lorentz force $\vec{F_L}$ produced by a velocity $\vec{u}$ and a DC magnetic field. With transverse (a~b) and axial (c~d) magnetic fields. The model in a and c is 2D transverse (see Fig.~\ref{fig:Cu Tip}).}
    \label{fig:Lorentz}
\end{figure}

Considering a copper tip elongated by the radiation pressure, with a velocity field $\mathbf{u}$ inside the material, we can predict the influence of a DC magnetic field by evaluating the Lorentz force $\mathbf{F_L}$. From the results of the hydrodynamical domain (Fig.~\ref{fig:HD Model}) it is realistic to consider $\mathbf{u}$ oriented outward close to the apex, and inward in the flat part of the tip, see Fig.~\ref{fig:Lorentz}. In case the magnetic field $\mathbf{B_{DC}}$ is transverse, the Lorentz force is oriented along $-u_y$ (resp. $-u_z$ in 3D) and will counteract elongation (Fig.~\ref{fig:Lorentz}, $a$, $b$). In the case $\mathbf{B_{DC}}$ is axial along $u_y$ (resp. $u_z$ in 3D), only the transverse velocity, $u_x$ (resp. $u_r$ in 3D), contributes to the Lorentz force, which the Lorentz force is oriented outwards, and counteracts the tip radius reduction (Fig.~\ref{fig:Lorentz}, $c$, $d$). 

\section{Results of the magnetohydrodynamic model\label{sec:mhd}}
We simulated the deformation of the liquid copper emitter with an initial height of $180~\mu m$ and a radius of $30~\mu m$, since no noticeable effects of the DC magnetic field were observed on the nanometer length scale, as the radiation pressure and surface tension force surpass the Lorentz force in the nanometer length scale. We first tested $|\mathbf{B_{DC}}| = 3~T$, but due to an excessive amount of computation time, we decided to impose a higher DC magnetic field $|\mathbf{B_{DC}}| = 30~T$. We simulated both horizontal and vertical directions for the magnetic field. However, the vertical direction creates a lower Lorentz force, and the simulation did not allow reaching a point where the tip behavior is strongly affected. 

\begin{figure}[htb]
    \centering
    \includegraphics[width=0.45\textwidth]{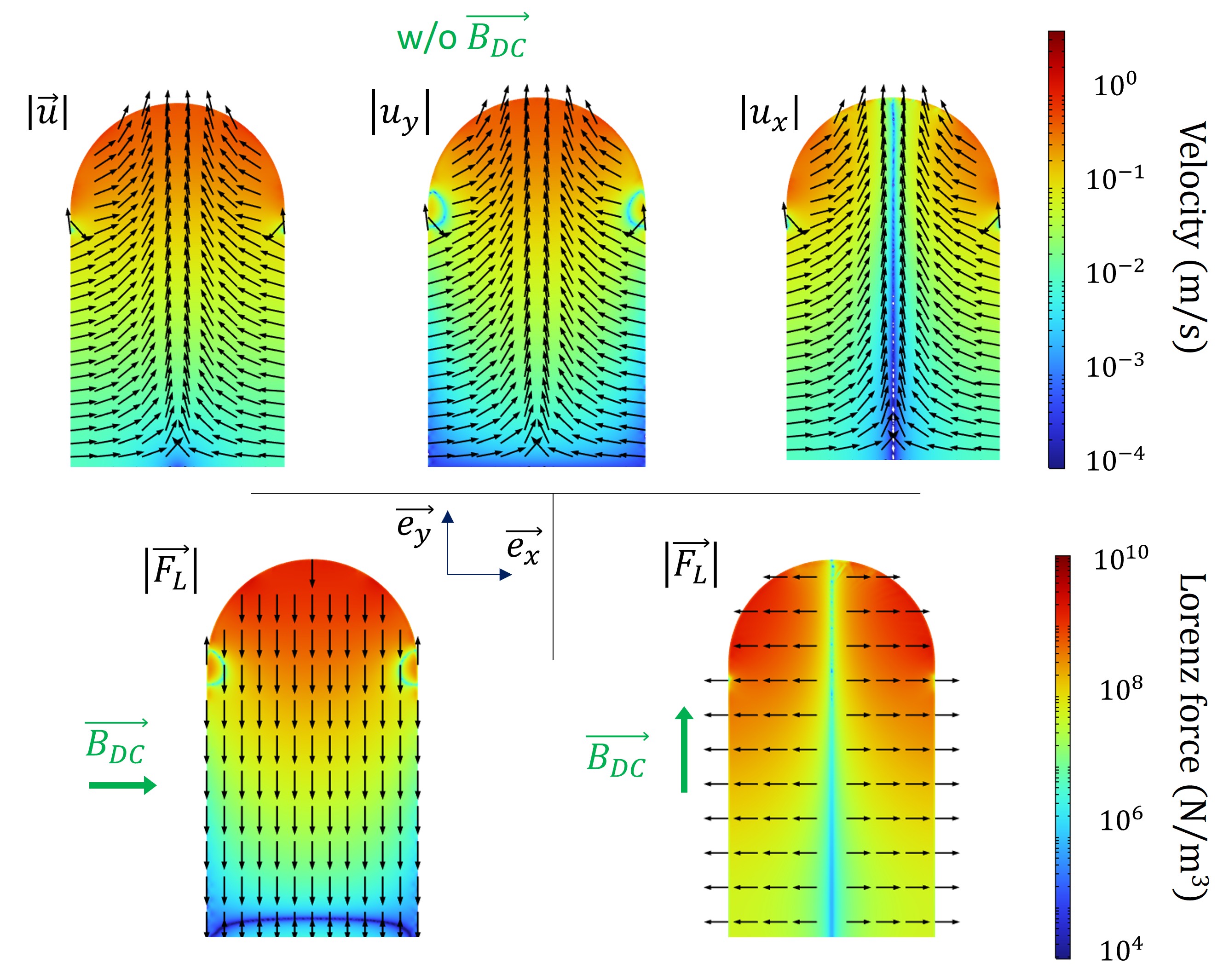}
    \caption{Top: velocity field $\vec{u}$ inside the copper tip (black arrows) in absence of magnetic field. Norm of $\vec{u_y}$, $\vec{u_x}$ (colormap). Bottom : Lorentz force (field and amplitude) for a transverse (left) and axial (right) magnetic field. Simulation at $t~=~5~10^{-7}s$.}
    \label{fig:v_FL}
\end{figure}

Figure~\ref{fig:v_FL} shows several field patterns at the very beginning of the simulation ($t~=~5~10^{-7}s$). At the top, the velocity field $\mathbf{u}$ is displayed in the case without magnetic field. The $u_y$ component is concentrated at the tip apex, while the $u_x$ component is concentrated on the tip sides. At the bottom, the simulated Lorentz force pattern is displayed in presence of $B_x$ and $B_y$, and is consistent with the prediction in Fig.~\ref{fig:Lorentz}.  

Finally, the deformation of the liquid copper is shown at $t~=~3~10^{-5}s$ without magnetic field and with an horizontal magnetic field.

 \begin{figure}[htb]
    \centering
    \includegraphics[width=0.45\textwidth]{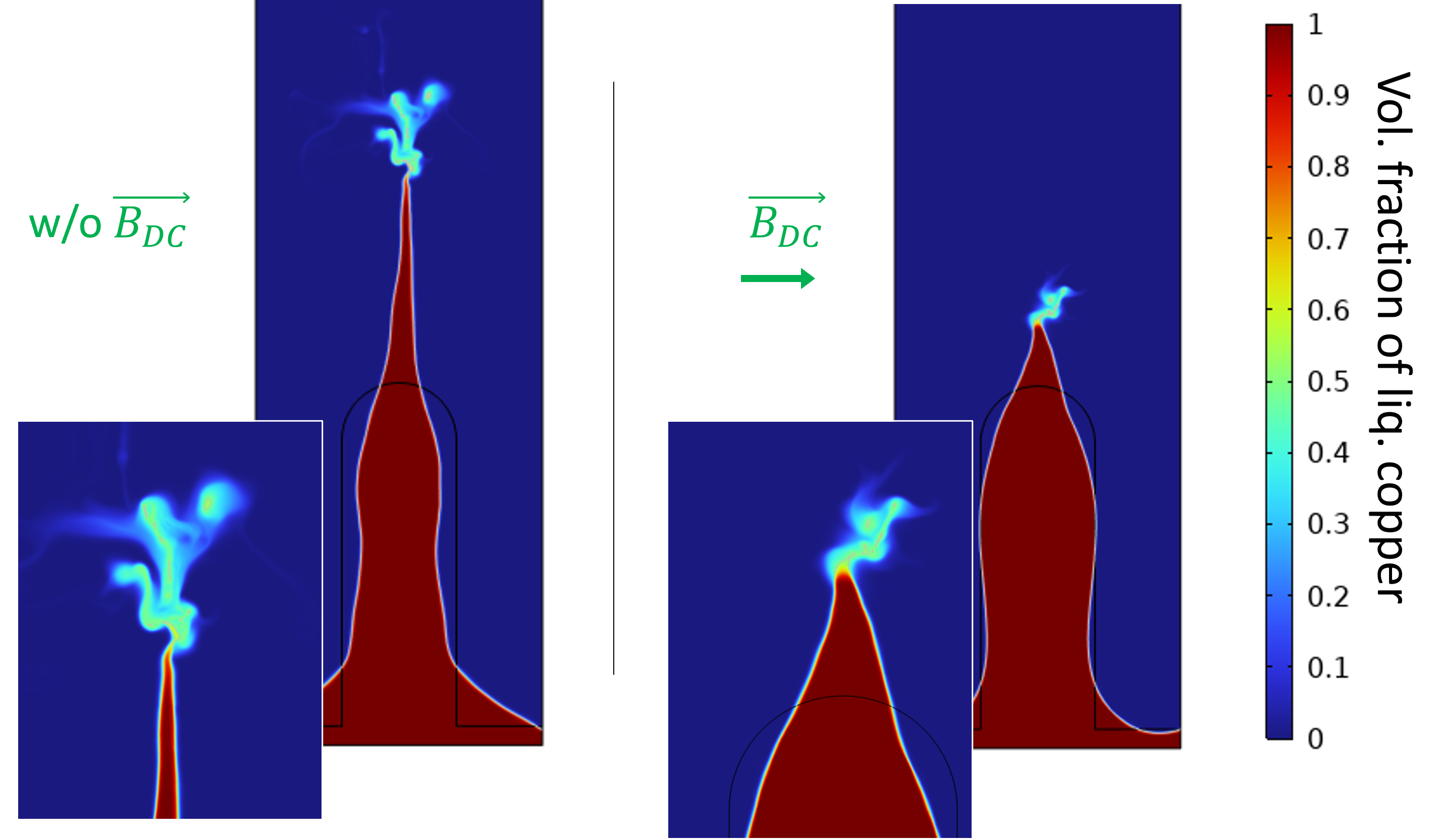}
    \caption{Comparison between an hydrodynamic model (left) and a magnetohydrodynamic model (right) for a liquid copper emitter of an initial height of $180~\mu m$ and a radius of $30~\mu m$. Volume fraction of liquid copper (colormap). Simulation at $t~=~3~10^{-5}s$. The full animation is available in the supplemental material \cite{animation}.}
    \label{fig:HD vs MHD}
\end{figure}

The Lorentz force acts as a friction force that slows the deformation process. Fig.~\ref{fig:HD vs MHD} compares the copper tip behavior between $|\mathbf{B_{DC}}|$~=~0 and $|\mathbf{B_{DC}}|$~=~30~T transverse  when the elongation becomes clear. It shows that the tip deformation is smoother in presence of the magnetic field. The results corroborate the prediction of the previous section.

The zoom on the tip apex shows that the dynamics of liquid copper ejection is probably affected. However, its dependence on $|\mathbf{B_{DC}}|$ is still unclear due to the numerical instability encountered when the radius of the apex becomes too small. 

In conclusion, at this stage of our study, the $|\mathbf{B_{DC}}|$ field effect seems to reduce the tip elongation phenomena, responsible for the liquid copper ejection. Finally, our results show no evidence that the RF breakdown phenomenon is amplified by the presence of a magnetic field. This will be discussed in Section \ref{sec:disc}.

\section{Comparison with a multi-physics atomistic model\label{sec:comp}}

Some models of a copper tip deformation by an electric field have been described in the literature. In particular, Kyritsakis et al. have presented a multi-physics atomistic (MPA) model based on the gradual deformation of an emitting tip under field-induced forces \cite{timko_field_2015,djurabekova_atomistic_2011,kyritsakis_thermal_2018}.
In this section, we compare this atomistic model, its hypothesis and results, with our hydrodynamic (HD) model.

In the MPA model, the forces are applied on virtual particles and described by Molecular Dynamics simulations, including the interatomic forces. The model is 3D, and considers a DC electric field. The local electric field distribution evolves according to the emitter shape, and the space charge effect is included. The Lorentz force on a charge $q_i$ under a local field $\vec{E_i}$ is $ \vec{F_i}=1/2~ q_i \vec{E_i}$. The heating due to the field emission is included, leading to thermal effects on the metal. The atomistic nature of the model do not discriminate between liquid or gas nature. The main differences between the two models are shown in Table~\ref{tab:diff}. 

\begin{table}
\begin{center}
\begin{tabular}{| c | c|} 
 \hline
 HD model & MPA model  \\ [0.5ex] 
 \hline \hline
 2 D & 3 D\\ 
 \hline
 RF field & DC field\\ 
 \hline
 w/o space charge & w space charge\\
 \hline
 constant temperature & thermal effects \\
 \hline
 liquid phase & gas and liquid phases \\
 [1ex] 
 \hline
\end{tabular}
\end{center}
\caption{Main differences between the hydrodynamical (HD) and atomistic (MPA) models.}
\label{tab:diff}
\end{table}

It is interesting to compare the forces experienced by the copper tip in both models. In the MPA model, atom interactions in a fluid are modelized by the Coulombic force and the interatomic potential \cite{djurabekova_atomistic_2011}. In the HD model, these interactions are separated in two components, the viscosity and the surface tension. The relationship between the surface tension (parameter: surface tension coefficient $\gamma_{st}$) and the Van der Waals forces applying on atoms (parameters: polarizability, ionization energy, Van der Waals radius) relies on the energy needed to move an atom from the inside to the surface of the tip \cite{Butt}. The viscosity (parameter: $\eta$) in liquids is governed by the activation energy
barrier for molecular rearrangements. However the viscosity microscopic mechanisms at the individual particle level remain a subject of ongoing research \cite{huang_2025}.

Due to the DC nature of the electric field in the MPA model, the resulting surface pressure at the tip interface is $1/2~ \epsilon_0  E_{MPA}^2$, compared to the the radiation surface pressure in the HD model (the time-averaged RF electromagnetic force \cite{Prad}): $1/4~ \epsilon_0  E_{HD}^2$. Both $E_{MPA}$ and $E_{HD}$ are the local fields at the tip surface. 

Now we propose a numerical comparison between the models, applied to a copper tip with initial radius of 3~nm, and initial height in the simulation domain equal to 46.6~nm (MPA, \cite{kyritsakis_thermal_2018}) and 18~nm (HD, Sec.~\ref{des_hd}). The theoretical threshold needed to pull the tip can be calculated in both cases (Eq.~\ref{eqn:threshold}). Since the local curvature for a tip with a radius R is $\kappa = 1/R$  for a 2D model and $\kappa = 2/R $ for a 3D model, the threshold field ratio between an 3D and a 2D model is $\sqrt{2}$. On the other hand, the surface pressure ratio between an RF and a DC  model is $1/2$, resulting in a $1/\sqrt{2}$ ratio between RF and a DC  models.
Combining these ratio, and naming $E_{MPA}^{th}$ and $E_{HD}^{th}$ the threshold local fields at the tip apex, we arrive to: 
\begin{equation}
    E_{MPA}^{th}~=~E_{HD}^{th}
\end{equation}

Now, using Eq.~\ref{eqn:threshold} with the surface tension copper coefficient $\gamma_{st}$~=~1.3~N/m and the radius $R$~=~3~nm, we arrive to:
\begin{equation}
    E_{MPA}^{th}~=~E_{HD}^{th}~=~14~GV/m
\end{equation}

\begin{figure}[htb]
    \centering
    \includegraphics[width=0.45\textwidth]{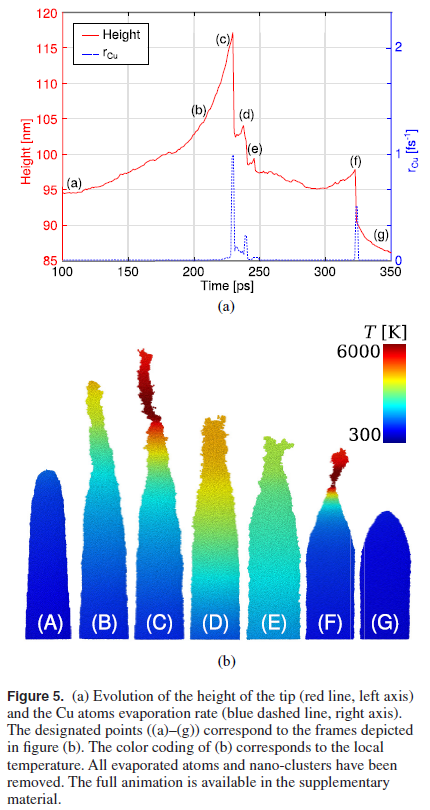}
    \caption{MD model: Figure taken from \cite{kyritsakis_thermal_2018} }
    \label{fig:krys}
\end{figure}

In the HD model, due to an additional downward force caused by the curvature of the fluid at its base, and the very long simulation time for lower values of the field, we had to apply an initial field equal to $E_{HD}$~=~42~GV/m at the tip apex to observe a clear stretching. The MPA model results show a tip elongation with $E_{MPA}$~=~17.5~GV/m \cite{kyritsakis_thermal_2018}. Both values have the same order of magnitude and are beyond the calculated threshold.

The comparison between Fig.~\ref{fig:Tip Elongation} and Fig.~\ref{fig:krys} show that the tip deformation process before particle emission has a similar behavior for HD and  for MPA, with a duration of tens of picoseconds in both cases.  

Finally, despite the differences shown in Table~\ref{tab:diff}, we observe similar results for the HD and MPA models in the first elongation process.

\section{Discussion\label{sec:disc}}

Our hydrodynamical model provides a new approach to understand the RF breakdown mechanism, and the preliminary results show that BD could be triggered by a sufficiently high local electric field. However, there are some limitations in our current models that need to be addressed. The variation in temperature due to Joule heating and Nottingham effect was not included in the current model. Since the viscosity of the  copper decreases with temperature rise, the liquid copper should deform more easily, and likely increasing the tip elongation and copper ejection. Moreover, if the temperature is sufficiently high, phase transition into gas or plasma will occur, and our hydrodynamical model will not be suitable to describe this phenomenon anymore. However, our comparison with the multi-physics atomistic shows good agreement in the first elongation phase.

The magnetohydrodynamics model has allow the evaluation of the Lorentz force influence on the copper tip fluid. We have shown how the addition of a DC magnetic field tends to slow down the deformation process, but the observed effects are small. This effect is not likely to explain the increase of the breakdown phenomenon observed in past experiments. Certainly, the copper tip modelisation at its apex should involve further physics phenomena, among them the field emission current and the plasma generation. The DC magnetic field effect on these phenomena should be studied. From the electric field calculated in our HD model, the field emission current can be evaluated from the Fowler-Nordheim formula. When the tip radius is 3~nm (Sec.~\ref{sec:hd}), the emitted current is in the $[10^{14}-10^{18}]~A/m^2$ range. The induced current, calculated with the copper velocity and conductivity (Eq.~\ref{eq:ohm}) is in the $[10^{9}-10^{11}]~A/m^2$ range. This might lead one to believe that the Lorentz force due to magnetic field effect $\mathbf{J} \times \mathbf{B}$ (Eq.~\ref{eqn:Lorentz}) dominates by the emitted current. Similarly, when the tip radius is 30~µm (Sec.~\ref{sec:mhd}), the emitted current is lower than $10^{7}~A/m^2$ until the very end of the simulation, while the induced current is in the $[10^{8}-10^{10}]~A/m^2$ range. In this case, the Lorentz force due to the magnetic field effect should be dominated by the induced current. However, the emitted current is mainly external to the copper tip, outside of the magnetohydrodynamic domain. It could be interesting to explore a very thin layer at the tip surface, where the emitted current could be taken into account in of this domain. Those considerations require more studies, but they suggest the necessity of evaluating the magnetic field effect on the emitted current, and its consequence on the behavior of the copper tip surrounding, including the presence of a plasma.

\section{Conclusions}
To conclude, we propose a model describing the starting of the BD mechanism based on some past observations. Our hydrodynamic model can predict a threshold value for a possible BD event to occur. We have compared this model with a multi-physics atomistic proposed by Kyritsakis et al.  with a good agreement in the first elongation phase. Our magnetohydrodynamic model tries to understand the effects of a DC magnetic field acting on the liquid field emitter. We found a reduction in the tip elongation phenomena, responsible for the liquid copper ejection. Finally, we propose further studies for the understanding of the magnetic field effect on the breakdown phenomena.

\begin{acknowledgments}
Funded by the European Union (EU). Views and opinions expressed are however those of the author(s) only and do not necessarily reflect those of the EU or European Research Executive Agency (REA). Neither the EU nor the REA can be held responsible for them.

Endorsed by the IMCC.
\end{acknowledgments}

\bibliography{References}

@techreport{Prad,
  author       = {E.Haebel and J.Tückmantel},
  title        = {Electromagnetic Surface Forces in RF Cavities},
  institution  = {CERN},
  year         = {1991},
  number       = {CM-P00058119},
  note         = {CERN Document Server}
}

@misc{tikhonin2008,
      title={On the effects at colliding mu-meson beams}, 
      author={F. F. Tikhonin},
      year={2008},
      eprint={0805.3961},
      archivePrefix={arXiv},
      primaryClass={hep-ph},
      howpublished={https://arxiv.org/abs/0805.3961}, 
}

@article{long_muon_2021,
	title = {Muon colliders to expand frontiers of particle physics},
	volume = {17},
	issn = {1745-2481},
	doi = {10.1038/s41567-020-01130-x},
	number = {3},
	journal = {Nature Physics},
	author = {Long, K. R. and Lucchesi, D. and Palmer, M. A. and Pastrone, N. and Schulte, D. and Shiltsev, V.},
	month = mar,
	year = {2021},
	pages = {289--292},
}

@misc{MuCol,
  howpublished        = {https://mucol.web.cern.ch/},
  author       = {MuCol},

}

@misc{IMCC,
  howpublished        = {https://muoncollider.web.cern.ch/},
  author       = {IMCC},
}

@article{Zhu_2025,
  title = {Design method, performance evaluation, and tolerance analysis of the rectilinear cooling channel for a muon collider},
  author = {Zhu, Ruihu and Rogers, Chris and Yang, Jiancheng and Zhao, He and Guo, Cheng and Li, Jiangdong},
  journal = {Phys. Rev. Accel. Beams},
  volume = {28},
  issue = {4},
  pages = {041003},
  numpages = {14},
  year = {2025},
  month = {Apr},
  publisher = {American Physical Society},
  doi = {10.1103/PhysRevAccelBeams.28.041003},
  url = {https://link.aps.org/doi/10.1103/PhysRevAccelBeams.28.041003}
}

@misc{ESPPU_2026,
      title={The International Muon Collider Collaboration},
      year={2026},
      howpublished={https://edms.cern.ch/document/3284682/1},
      }

@article{nanni_2025,
	author = {Nanni, Emilio},
	title = {Update on RF test facilities},
	howpublished = {https://indico.desy.de/event/45968/contributions/186253/},
	journal = {IMCC and MuCol annual meeting 2025},
	year = {2025},
}

@misc{plouin_2025,
	author = {Plouin, Juliette},
	title = {Review and discussion on RF Gradient in strong magnetic field},
    howpublished = {https://indico.cern.ch/event/1566248/},
	year = {2025},
	journal = {IMCC meetings},
}

@article{MICE_2020,
	title = {Demonstration of cooling by the Muon Ionization Cooling Experiment},
	volume = {578},
	issn = {1476-4687},
	url = {https://doi.org/10.1038/s41586-020-1958-9},
	doi = {10.1038/s41586-020-1958-9},
	number = {3},
	journal = {Nature},
	author = {{MICE collaboration}},
	month = feb,
	year = {2020},
	pages = {53–59},
}

@techreport{Palmer_2013,
    author = "Palmer, M. A.",
    institution = "Fermilab",
    collaboration = "Neutrino Factory, Muon Collider",
    title = "{An Overview of the US Muon Accelerator Program}",
    booktitle = "{International Workshop on Beam Cooling and Related Topics}",
    reportNumber = "FERMILAB-CONF-13-253-APC",
    month = "6",
    year = "2013"
}

@misc{wuensch_fundamentals_2025,
	title = {Fundamentals of {Vacuum} {Breakdown} in {High}-{Field} {Systems}},
	url = {http://arxiv.org/abs/2502.03967},
	doi = {10.48550/arXiv.2502.03967},
		urldate = {2025-11-14},
	publisher = {arXiv},
	author = {Wuensch, Walter and Calatroni, Sergio and Djurabekova, Flyura and Kyritsakis, Andreas and Ashkenazy, Yinon},
	month = nov,
	year = {2025},
	note = {arXiv:2502.03967 [physics]},
}

@article{antoine_erratum_2012,
	title = {Erratum to: {Electromigration} occurences and its effects on metallic surfaces submitted to high electromagnetic field: {A} novel approach to breakdown in accelerators},
	volume = {670},
	copyright = {https://www.elsevier.com/tdm/userlicense/1.0/},
	issn = {01689002},
	shorttitle = {Erratum to},
	url = {https://linkinghub.elsevier.com/retrieve/pii/S0168900212000678},
	doi = {10.1016/j.nima.2012.01.027},
	urldate = {2025-03-27},
	journal = {Nuclear Instruments and Methods in Physics Research Section A: Accelerators, Spectrometers, Detectors and Associated Equipment},
	author = {Antoine, C.Z. and Peauger, F. and Le Pimpec, F.},
	month = apr,
	year = {2012},
	pages = {79--94},
}

@misc{abe_2019,
	title = {Minuscule {Gremlins} {Cause} {Vacuum} {Breakdown} in {Radio}-{Frequency} {Accelerating} {Cavities}},
	howpublished = {https://www2.kek.jp/accl/eng/topics/topics190122.html},
	urldate = {2025-03-27},
	journal = {Accelerator Laboratory - KEK ACCL},
	author = {Abe, Tetsuo},
	month = jan,
	year = {2019},
}

@article{abe_2018,
  title = {Direct observation of breakdown trigger seeds in a normal-conducting rf accelerating cavity},
  author = {Abe, Tetsuo and Kageyama, Tatsuya and Sakai, Hiroshi and Takeuchi, Yasunao and Yoshino, Kazuo},
  journal = {Phys. Rev. Accel. Beams},
  volume = {21},
  issue = {12},
  pages = {122002},
  numpages = {15},
  year = {2018},
  month = {Dec},
  publisher = {American Physical Society},
  doi = {10.1103/PhysRevAccelBeams.21.122002},
  url = {https://link.aps.org/doi/10.1103/PhysRevAccelBeams.21.122002}
}

@inproceedings{abe_2025,
    author = "Abe, Tetsuo",
    title = "{Fireball Hypothesis for the Trigger of Sudden Beam Losses AT SuperKEKB}",
    booktitle = "{20th Annual Meeting of Particle Accelerator Society of Japan}",
    year = {2025}
}

@article{palmer_rf_2009,
	title = {rf breakdown with external magnetic fields in 201 and 805 {MHz} cavities},
	volume = {12},
	url = {https://link.aps.org/doi/10.1103/PhysRevSTAB.12.031002},
	doi = {10.1103/PhysRevSTAB.12.031002},
		number = {3},
	urldate = {2025-03-07},
	journal = {Physical Review Special Topics - Accelerators and Beams},
	author = {Palmer, R. B. and Fernow, R. C. and Gallardo, Juan C. and Stratakis, Diktys and Li, Derun},
	month = {March},
	year = {2009},
	note = {Publisher: American Physical Society},
	pages = {031002},
}

@article{moretti_effects_2005,
	title = {Effects of high solenoidal magnetic fields on rf accelerating cavities},
	volume = {8},
	copyright = {http://creativecommons.org/licenses/by/3.0/},
	issn = {1098-4402},
	url = {https://link.aps.org/doi/10.1103/PhysRevSTAB.8.072001},
	doi = {10.1103/PhysRevSTAB.8.072001},
	number = {7},
	urldate = {2025-03-07},
	journal = {Physical Review Special Topics - Accelerators and Beams},
	author = {Moretti, A. and Qian, Z. and Norem, J. and Torun, Y. and Li, D. and Zisman, M.},
	month = jul,
	year = {2005},
	pages = {072001},
}

@article{bowring_operation_2020,
	title = {Operation of normal-conducting rf cavities in multi-{Tesla} magnetic fields for muon ionization cooling: {A} feasibility demonstration},
	volume = {23},
	issn = {2469-9888},
	shorttitle = {Operation of normal-conducting rf cavities in multi-{Tesla} magnetic fields for muon ionization cooling},
	url = {https://link.aps.org/doi/10.1103/PhysRevAccelBeams.23.072001},
	doi = {10.1103/PhysRevAccelBeams.23.072001},
	number = {7},
	urldate = {2025-03-07},
	journal = {Physical Review Accelerators and Beams},
	author = {Bowring, D. and Bross, A. and Lane, P. and Leonova, M. and Moretti, A. and Neuffer, D. and Pasquinelli, R. and Peterson, D. and Popovic, M. and Stratakis, D. and Yonehara, K. and Kochemirovskiy, A. and Torun, Y. and Adolphsen, C. and Ge, L. and Haase, A. and Li, Z. and Martin, D. and Chung, M. and Li, D. and Luo, T. and Freemire, B. and Liu, A. and Palmer, M.},
	month = jul,
	year = {2020},
	pages = {072001},
}

@article{norem_dark_2003,
	title = {Dark current, breakdown, and magnetic field effects in a multicell, 805 {MHz} cavity},
	volume = {6},
	copyright = {http://creativecommons.org/licenses/by/3.0/},
	issn = {1098-4402},
	url = {https://link.aps.org/doi/10.1103/PhysRevSTAB.6.072001},
	doi = {10.1103/PhysRevSTAB.6.072001},
	number = {7},
	urldate = {2025-03-21},
	journal = {Physical Review Special Topics - Accelerators and Beams},
	author = {Norem, J. and Wu, V. and Moretti, A. and Popovic, M. and Qian, Z. and Ducas, L. and Torun, Y. and Solomey, N.},
	month = jul,
	year = {2003},
	pages = {072001},
}

@article{kyritsakis_thermal_2018,
	title = {Thermal runaway of metal nano-tips during intense electron emission},
	volume = {51},
	issn = {0022-3727, 1361-6463},
	url = {http://arxiv.org/abs/1710.00050},
	doi = {10.1088/1361-6463/aac03b},
	number = {22},
	urldate = {2025-03-31},
	journal = {Journal of Physics D: Applied Physics},
	author = {Kyritsakis, A. and Veske, M. and Eimre, K. and Zadin, V. and Djurabekova, F.},
	month = jun,
	year = {2018},
	note = {arXiv:1710.00050 [cond-mat]},
	pages = {225203},
}

@article{djurabekova_atomistic_2011,
	title = {Atomistic modeling of metal surfaces under electric fields: {Direct} coupling of electric fields to a molecular dynamics algorithm},
	volume = {83},
	copyright = {http://link.aps.org/licenses/aps-default-license},
	issn = {1539-3755, 1550-2376},
	shorttitle = {Atomistic modeling of metal surfaces under electric fields},
	url = {https://link.aps.org/doi/10.1103/PhysRevE.83.026704},
	doi = {10.1103/PhysRevE.83.026704},
	number = {2},
	urldate = {2025-09-23},
	journal = {Physical Review E},
	author = {Djurabekova, F. and Parviainen, S. and Pohjonen, A. and Nordlund, K.},
	month = feb,
	year = {2011},
	pages = {026704},
}

@article{timko_field_2015,
	title = {From {Field} {Emission} to {Vacuum} {Arc} {Ignition}: {A} {New} {Tool} for {Simulating} {Copper} {Vacuum} {Arcs}},
	volume = {55},
	issn = {1521-3986},
	shorttitle = {From {Field} {Emission} to {Vacuum} {Arc} {Ignition}},
	doi = {10.1002/ctpp.201400069},
	number = {4},
	journal = {Contributions to Plasma Physics},
	author = {Timko, H. and Ness Sjobak, K. and Mether, L. and Calatroni, S. and Djurabekova, F. and Matyash, K. and Nordlund, K. and Schneider, R. and Wuensch, W.},
	year = {2015},
	pages = {299--314},
}

@article{alfven_1942,
	title = {Existence of {Electromagnetic}-{Hydrodynamic} {Waves}},
	volume = {150},
	copyright = {1942 Springer Nature Limited},
	issn = {1476-4687},
	url = {https://www.nature.com/articles/150405d0},
	doi = {10.1038/150405d0},
	number = {3805},
	urldate = {2026-08-07},
	journal = {Nature},
	author = {Alfvén, H.},
	month = oct,
	year = {1942},
	note = {Publisher: Nature Publishing Group},
	pages = {405--406},
}

@book{davidson2017,
author = {Davidson, P. A.},
address = {Cambridge},
edition = {Second edition.},
isbn = {9781316672853},
publisher = {Cambridge University Press},
series = {Cambridge texts in applied mathematics},
title = {Introduction to magnetohydrodynamics},
year = {2017},
}

@article{taylor_disintegration_1964,
	title = {Disintegration of water drops in an electric field},
	volume = {280},
	issn = {0080-4630},
	doi = {10.1098/rspa.1964.0151},
	number = {1382},
	urldate = {2026-08-07},
	journal = {Proceedings of the Royal Society of London. A. Mathematical and Physical Sciences},
	author = {Taylor, Geoffrey Ingram},
	month = jul,
	year = {1964},
	pages = {383--397},
}

@inbook{Butt,
publisher = {John Wiley and Sons, Ltd},
isbn = {9783527602315},
title = {Liquid Surfaces},
booktitle = {Physics and Chemistry of Interfaces},
author = {Hans-Jürgen Butt, H-J. and Graf, K. and Kappl, M.},
chapter = {2},
pages = {4-25},
doi = {10.1002/3527602313.ch2},
year = {2003}
}

@article{huang_2025,
	title = {Atomistic mechanisms of viscosity in 2D liquid-like fluids},
	volume = {16},
	issn = {2041-1723},
	doi = {10.1038/s41467-025-65246-5},
	number = {10171},
	journal = {Nature Communications},
	author = {Huang, Dong and Liang, Chen and Baggioli, Matteo and Feng, Yan},
	month = nov,
	year = {2025},
}

@misc{comsol1,
  url        = {https://www.comsol.com/blogs/two-methods-for-modeling-free-surfaces-in-comsol-multiphysics},
  author = {COMSOL},
  year   = {2018},
  title       = {Two Methods for Modeling Free Surfaces in COMSOL Multiphysics},
}

@misc{comsol2,
  url        ={https://doc.comsol.com/6.4/docserver/#!/com.comsol.help.cfd/cfd_ug_fluidflow_multi.09.128.html},
  author = {COMSOL},
  title       = {Level Set and Phase Field Equations},
}

@misc{mhd_ensta,
    author = {Boisson, Jean} ,
    url = {https://perso.ensta.fr/~boisson/cours_MHD/Poly/Intro_a_la_MHD.pdf},
    title = {Introduction à la Magnétohydrodynamique},
    year = {2020},
    type = {University lecture}
}

@misc{animation,
    url = {https://perso.ensta.fr/~boisson/cours_MHD/Poly/Intro_a_la_MHD.pdf},
    title = {See Supplemental Material at [URL will be inserted by publisher] for animation of  hydrodynamic and a magnetohydrodynamic models for a liquid copper emitter of an initial height of $180~\mu m$ and a radius of $30~\mu m$.},

}

\end{document}